\documentclass[conference,10pt,letter]{IEEEtran}
\usepackage{amsmath,amssymb,graphicx}
\usepackage{url}
\usepackage{circuitikz}
\usepackage{pgfplots}
\usetikzlibrary{calc}
\usepackage{balance}

\title{Non-Linear Digital Self-Interference Cancellation for In-Band Full-Duplex Radios Using\\ Neural Networks}
\author{\IEEEauthorblockN{Alexios Balatsoukas-Stimming}
\IEEEauthorblockA{Telecommunications Circuits Laboratory\\{\'E}cole polytechnique f{\'e}d{\'e}rale de Lausanne, CH-1015 Lausanne, Switzerland\\alexios.balatsoukas@epfl.ch}
}
\begin{document}
%
\maketitle
\begin{abstract}
Full-duplex systems require very strong self-interference cancellation in order to operate correctly and a significant part of the self-interference signal is due to non-linear effects created by various transceiver impairments. As such, linear cancellation alone is usually not sufficient and sophisticated non-linear cancellation algorithms have been proposed in the literature. In this work, we investigate the use of a neural network as an alternative to the traditional non-linear cancellation method that is based on polynomial basis functions. Measurement results from a full-duplex testbed demonstrate that a small and simple feed-forward neural network canceler works exceptionally well, as it can match the performance of the polynomial non-linear canceler with significantly lower computational complexity.
\end{abstract}
%

\section{Introduction}
In-band full-duplex (FD)~\cite{Jain2011,Duarte2012,Bharadia2013} is a promising method to increase the spectral efficiency of current communications systems by transmitting and receiving data simultaneously in the same frequency band. In order for an FD node to operate correctly, the strong self-interference (SI) signal that is produced at the node's receiver by its own transmitter needs to be effectively canceled. 

A combination of SI cancellation in the analog and in the the digital domain is usually necessary in order to suppress the SI signal down to the level of the receiver noise floor. Analog cancellation can be either passive (i.e., through physical isolation between the transmitter and the receiver) or active (i.e., through the injection of a cancellation signal) and it is necessary in order to avoid saturating the analog front-end of the receiver. However, perfect cancellation in the analog domain is very challenging and costly to achieve, meaning that a residual SI signal is still present at the receiver after the analog cancellation stage. In principle, this residual SI signal should be easily cancelable in the digital domain, since it is caused by a signal that is fully known. Unfortunately, in practice this is not the case as several transceiver non-linearities, such as various baseband non-linearities (e.g., digital-to-analog converter (DAC) and analog-to-digital converter (ADC))~\cite{Balatsoukas2015}, IQ imbalance~\cite{Balatsoukas2015,Korpi2014}, phase-noise~\cite{Sahai2013,Syrjala2014}, and power amplifier (PA) non-linearities~\cite{Balatsoukas2015,Korpi2014,Anttila2014,Korpi2017}, distort the SI signal. This means that complicated non-linear cancellation methods are required in order to fully suppress the SI to the level of the receiver noise floor. These methods are based on polynomial expansions and the most recent and comprehensive model was presented in~\cite{Korpi2017}, where a parallel Hammerstein model was used for digital SI cancellation that incorporates both PA non-linearities and IQ imbalance. Polynomial models have been shown to work well in practice, but they can also have a high implementation complexity as the number of estimated parameters grows rapidly with the maximum considered non-linearity order and  because a large number of non-linear basis functions have to be computed. An effective complexity reduction technique that identifies the most significant non-linearity terms using principal component analysis (PCA) was also presented in~\cite{Korpi2017}. However, with this method the transmitted digital baseband samples need to be multiplied with a transformation matrix to generate the cancellation signal, thus introducing additional complexity. Moreover, as the authors mention, whenever the self-interference channel changes significantly, the PCA operation needs to be re-run.

\emph{Contribution:} In this work, we propose a non-linear SI cancellation method that uses a neural network to construct the non-linear part of the digital cancellation signal, as an alternative to the standard polynomial models that are used in the literature. Our initial experimental results using measured samples from a hardware testbed demonstrate that a simple neural network based non-linear canceler can already match the performance of a state-of-the-art polynomial model for non-linear cancellation with the same number of learnable parameters, but with a significantly lower computational complexity for the inference step (i.e., after training has been performed). Specifically, the neural network based non-linear canceler requires 36\% fewer real multiplications to be implemented and it does not require the computation of any non-linear basis function.

\emph{Related Work:} Over the years, there has been significant interest in the application of neural networks to various communications scenarios, which has been renewed recently with a particular focus on the physical layer~\cite{OShea2017}. As we are not aware of any applications of neural networks for SI cancellation in full-duplex radios in the literature, we briefly outline some other physical layer communications areas where neural networks have been successfully applied. In~\cite{OShea2017,OShea2016}, the entire transceiver, including the transmission channel and transceiver non-idealities, was treated as an auto-encoder neural network which can, in some cases, learn an end-to-end signal processing algorithm that results in better error rate performance than traditional signal processing algorithms. Detection for molecular communications using neural networks was considered in~\cite{Farsad2017}. The work of~\cite{Nachmani2016} considered a modification of the well-known belief propagation (BP) decoding algorithm for LDPC codes where weights are assigned to each message in the Tanner graph of the code that is being decoded and deep learning techniques are used in order to learn good values for these weights. A similar approach was taken in~\cite{Lugosch2017}, where the offset parameter of the offset min-sum (OMS) decoding algorithm are learned by using deep learning techniques. The work of~\cite{Gruber2017}~considered using a neural network in order to decode polar codes. In~\cite{Aazhang1992,Yang2000,Geevarghese2013}, neural networks were employed in order to perform detection and intra-user (and mostly linear) successive interference cancellation in multi-user CDMA systems. Finally, the work of~\cite{Sun2017} considered using neural networks for wireless resource management.

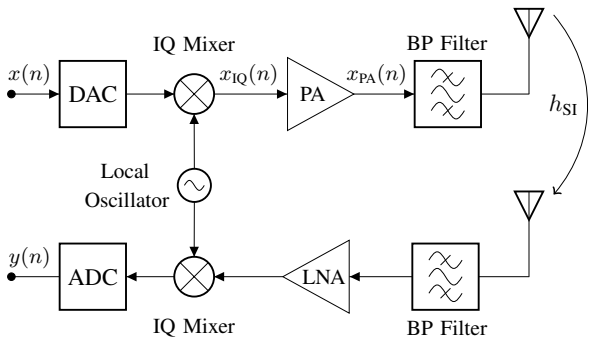
\begin{figure}
	\centering
	\scalebox{0.9}{\begin{circuitikz}[scale=0.9]

	
	\draw (3,0) node[mixer,scale=0.6] (txmixer) {};
	\draw (8.5,0) node[antenna,scale=0.6] (txantenna) {};
	
	\draw (0,0) to[short,*-] ++(0,0) to[twoport,>,t=DAC] (txmixer.west) node[inputarrow]{};
	\draw (txmixer.east) to[short,-] ++(1,0) to[amp,>,t=\small{PA}] ++(1.5,0) to[bandpass,>] (txantenna);
	
	\draw (0.3,0) node[above] {\small $x(n)$};
	\draw (txmixer)+(0,0.5) node[above] {\small{IQ Mixer}};
	\draw (txmixer)+(0.9,0) node[above] {\small $x_{\text{IQ}}(n)$};
	\draw (txantenna)+(-1.35,0.575) node[above] {\small{BP Filter}};
	\draw (txantenna)+(-2.5,0) node[above] {\small $x_{\text{PA}}(n)$};
	
	\draw (8.5,-3) node[antenna,scale=0.6] (rxantenna) {};
	\draw (3,-3) node[mixer,scale=0.6] (rxmixer) {};
	
	\draw (rxantenna) to[short,-] ++(-0,0) to[bandpass] ++(-2.75,0) to[amp,>,t={\rotatebox[origin=c]{180}{\small{LNA}}}] ++(-1.5,0) to[short,-] (rxmixer.east) node[inputarrow,rotate=180]{};
	\draw (rxmixer.west) to[twoport,>,t=ADC] (0,-3) to[short,-*] (0,-3);
	\draw [->] (txantenna)+(0.4,1.3) to[thick, out=-50, in=50, edge node={node [left] {$h_{\text{SI}}$}}]  ($(rxantenna) + (0.4,1.35)$);
	
	\draw (rxantenna)+(-1.35,-0.575) node[below] {\small{BP Filter}};
	\draw (0.3,-3) node[above] {\small $y(n)$};
	\draw (rxmixer)+(0,-0.55) node[below] {\small{IQ Mixer}};
	
	\draw (3.275,-1.5) node[oscillator,scale=0.5] (ref) {};
	\draw (ref.south) to[short,-] (rxmixer.north) node[inputarrow,rotate=270]{};
	\draw (ref.north) to[short,-] (txmixer.south) node[inputarrow,rotate=90]{};
	\draw (ref)+(-0.6,0) node[left,text width=1.2cm,align=center] {\small{Local\\Oscillator}};
	
\end{circuitikz}}
	\caption{Basic model of a full-duplex transceiver with a shared local oscillator where some components have been omitted for simplicity. A more detailed model can be found in, e.g.,~\cite{Korpi2017}.}\label{fig:system}
	\vspace{-0.2cm}
\end{figure}

\section{Polynomial Non-Linear Canceler}\label{sec:polynomial}
In this section we briefly review a state-of-the-art polynomial model for non-linear digital cancellation that can mitigate the effects of both IQ imbalance and PA non-linearities~\cite{Anttila2014,Korpi2017}, which are usually the dominant non-idealities, while the remaining transceiver components are assumed to be ideal. This model will serve as the baseline for our comparison in Section~\ref{sec:results}. In Fig.~\ref{fig:system} we show a simple full-duplex transceiver architecture with a shared local oscillator, which is useful for the description of the polynomial non-linear cancellation model that follows.

Let the complex digital transmitted signal at time instant $n$ be denoted by $x(n)$. This digital signal is first converted to an analog signal by the digital-to-analog converter (DAC) and then upconverted by an IQ mixer. The digital baseband equivalent of the signal after the IQ imbalance introduced by the IQ mixer and assuming that the DAC is ideal can be modeled as~\cite{Korpi2017}
\begin{align}
	x_{\text{IQ}}(n)	& = K_1x(n) + K_2x^*(n), \label{eq:iq}
\end{align}
where $K_1, K_2 \in \mathbb{R}$ and typically $K_1 \gg K_2$. The output signal of the mixer is amplified by the PA, which introduces further non-linearities that can be modeled using a parallel Hammerstein model as~\cite{Korpi2017}
\begin{align}
	x_{\text{PA}}(n)	& = \sum _{\substack{p=1,\\p \text{ odd}}}^P \sum_{m=0}^M h_{\text{PA},p}(m)x_{\text{IQ}}(n-m)|x_{\text{IQ}}(n-m)|^{p-1}, \label{eq:pa}
\end{align}
where $h_{\text{PA},p}$ is the impulse response for the $p$-th order non-linearity and $M$ is the memory length of the PA. The $x_{\text{PA}}$ SI signal arrives at the receiver through an SI channel with impulse response $h_{\text{SI}}(l),~l=0,1,\hdots,(L-1)$. Assuming that the ADC and potential baseband amplifiers are ideal, the downconverted and digitized received SI signal $y(n)$ can be modeled as
\begin{align}
	y(n)	& = \sum_{l=0}^{L-1} h_{\text{SI}}(l)x_{\text{PA}}(n-l). \label{eq:SI}
\end{align}
By substituting \eqref{eq:iq} and \eqref{eq:pa} in \eqref{eq:SI} and performing some arithmetic manipulations~\cite{Anttila2014,Korpi2017}, $y(n)$ can be re-written as
\begin{align}
	y(n)	& = \sum _{\substack{p=1,\\p \text{ odd}}}^P \sum_{q=0}^p\sum_{m=0}^{M+L-1}h_{p,q}(m) x(n-m)^{q}x^*(n-m)^{p-q}, \label{eq:final}
\end{align}
where $h_{p,q}(m)$ is a channel containing the combined effects of $K_1$, $K_2$, $h_{\text{PA},p},$ and $h_{\text{SI}}$.

By adapting the expression of~\cite[Eq. (19)]{Korpi2017} to the case of a single antenna, we can calculate the total number of complex parameters $h_{p,q}(m)$ as
\begin{align}
	n_{\text{poly}}	& = (M+L)\left(\frac{P+1}{2}\right)\left(\frac{P+1}{2} + 1\right), \label{eq:nparam}
\end{align}
which grows quadratically with the PA non-linearity order $P$. The task of the non-linear digital canceler is to compute estimates of all $h_{p,q}$, which we denote by $\hat{h}_{p,q}$, and then construct an estimate of the SI signal, which we denote by $\hat{y}(n)$, using~\eqref{eq:final} and subtract it from the received signal in the digital domain. The amount of SI cancellation over a window of length $N$, expressed in dB, is
\begin{align}
	C_{\text{dB}}	& = 10\log_{10}\left( \frac{\sum _{n=0}^{N-1}|y(n)|^2}{\sum _{n=0}^{N-1}|y(n) - \hat{y}(n)|^2}\right).
\end{align}

\section{Neural Network Non-Linear Canceler}\label{sec:nn}
In this section, we first provide a brief background on neural networks and then we describe our proposed neural network based non-linear cancellation method.

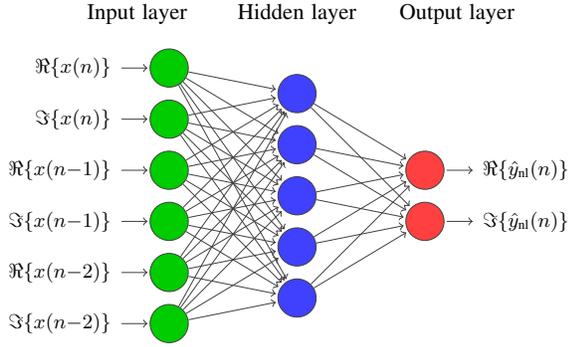
\begin{figure}
	\centering
	\scalebox{0.85}{\def\layersep{2.5cm}
\def\twolayersep{5cm}
\def\inputnodes{6}
\def\hiddennodes{5}
\def\outputnodes{2}

\begin{tikzpicture}[shorten >=1pt,->,draw=black!75, node distance=\layersep, scale=0.8]
    \tikzstyle{every pin edge}=[<-,shorten <=1pt]
    \tikzstyle{neuron}=[circle,draw=black!75,fill=black!75,minimum size=17pt,inner sep=0pt]
    \tikzstyle{input neuron}=[neuron, fill=black!20!green];
    \tikzstyle{output neuron}=[neuron, fill=red!75];
    \tikzstyle{hidden neuron}=[neuron, fill=blue!75];
    \tikzstyle{annot} = [text width=10em, text centered]
	
	\node[input neuron, pin=left: {\small$\Re{\left\{x(n)\right\}}$}] (I-1) at (0,-1) {};
	\node[input neuron, pin=left: {\small$\Im{\left\{x(n)\right\}}$}] (I-2) at (0,-2) {};
	\node[input neuron, pin=left: {\small$\Re{\left\{x(n{-}1)\right\}}$}] (I-3) at (0,-3) {};
	\node[input neuron, pin=left: {\small$\Im{\left\{x(n{-}1)\right\}}$}] (I-4) at (0,-4) {};
	\node[input neuron, pin=left: {\small$\Re{\left\{x(n{-}2)\right\}}$}] (I-5) at (0,-5) {};
	\node[input neuron, pin=left: {\small$\Im{\left\{x(n{-}2)\right\}}$}] (I-6) at (0,-6) {};

    \foreach \name / \y in {1,...,\hiddennodes}
        \path[yshift=-0.5cm]
            node[hidden neuron] (H-\name) at (\layersep,-\y cm) {};

	\foreach \name / \y in {1,...,\outputnodes}
		\path[yshift=-2cm]
			node[output neuron,pin={[pin edge={->}]right:\ifodd\y {\small $\Re{\left\{\hat{y}_{\text{nl}}(n)\right\}}$} \else {\small $\Im{\left\{\hat{y}_{\text{nl}}(n)\right\}}$} \fi}] (O-\name) at (\twolayersep,-\y) {};

    \foreach \source in {1,...,\inputnodes}
        \foreach \dest in {1,...,\hiddennodes}
			\path (I-\source) edge (H-\dest);

    \foreach \source in {1,...,\hiddennodes}
		\foreach \dest in {1,...,\outputnodes}
			\path (H-\source) edge (O-\dest);

    \node[annot,above of=H-1, node distance=1.25cm] (hl) {Hidden layer};
    \node[annot,left of=hl] {Input layer};
    \node[annot,right of=hl] {Output layer};
\end{tikzpicture}
}
	\caption{A neural network with 6 input, 5 hidden, and 2 output nodes.}\label{fig:nn}
	\vspace{-0.2cm}
\end{figure}

\subsection{Feed-forward Neural Networks}
Feed-forward neural networks are directed graphs that contain three types of nodes, namely input nodes, hidden nodes, and output nodes, which are organized in layers. An example of a feed-forward neural network with $6$ input nodes, $5$ hidden nodes, and $2$ output nodes is depicted in Fig.~\ref{fig:nn}. Each edge in the graph is associated with a weight. The input to each node of the graph is a weighted sum of the outputs of nodes in the previous layer, while the output of each node is obtained by applying a non-linear \emph{activation} function to its input.

The weights can be optimized through supervised learning by using training samples that contain known inputs and corresponding expected outputs. To this end, a cost function is associated with the output nodes, which measures the distance between the outputs of the neural network using the current weights and the expected outputs. The derivative of the cost function with respect to each of the weights in the neural network can be efficiently computed using back-propagation and it can then used in order to minimize the cost function using some gradient descent variant. Training is performed by splitting the data into mini-batches and performing a gradient descent update after processing each mini-batch. One pass through the entire training set is called a training \emph{epoch}.

\subsection{Neural Network Non-Linear Canceler}
The SI signal of \eqref{eq:final} can be decomposed as
\begin{align}
	{y}(n)	& = {y}_{\text{lin}}(n) + {y}_{\text{nl}}(n),
\end{align}
where ${y}_{\text{lin}}(n)$ is the linear part of~\eqref{eq:final} (i.e., the term of the sum with $p=1$ and $q=1$) and ${y}_{\text{nl}}(n)$ contains all remaining (non-linear) terms. We propose to use standard linear cancellation to construct an estimate of ${y}_{\text{lin}}(n)$, denoted by $\hat{y}_{\text{lin}}(n)$, while considering the much weaker ${y}_{\text{nl}}(n)$ signal as noise, and then reconstruct $y_{\text{nl}}(n)$ using a neural network. Specifically, the linear canceler first computes $\hat{h}_{1,1}$ using standard least-squares channel estimation~\cite{Anttila2014,Korpi2017}, and then uses $\hat{h}_{1,1}$ to construct $\hat{y}_{\text{lin}}(n)$ as follows
\begin{align}
	\hat{y}_{\text{lin}}(n) & = \sum_{m=0}^{M+L-1}\hat{h}_{1,1}(m) x(n-m).
\end{align}
The linear cancellation signal is then subtracted from the SI signal in order to obtain
\begin{align}
	{y}_{\text{nl}}(n)	& \approx y(n) - \hat{y}_{\text{lin}}(n),
\end{align}
The goal of the neural network is to reconstruct each ${y}_{\text{nl}}(n)$ sample based on the subset of $x$ that this ${y}_{\text{nl}}(n)$ sample depends on (cf. \eqref{eq:final}). Since neural networks generally operate on real numbers, we split all complex baseband signals into their real and imaginary parts. We note that, in principle the neural network could learn to cancel both the linear and the non-linear part of the signal. However, because the non-linear part of the SI signal is significantly weaker than the linear part, in practice our experiments indicate that the noise in the gradient computation due to the use of mini-batches essentially hides the non-linear structure from the learning algorithm.

We use a single layer feed-forward neural network as depicted in Fig.~\ref{fig:nn}. The neural network has $2(L+M)$ inputs nodes, which correspond to the real and imaginary parts of the $(M+L)$ delayed versions of $x$ in~\eqref{eq:final}, and two output nodes, which correspond to the real and imaginary parts of the target ${y}_{\text{nl}}(n)$ sample. The number of hidden nodes is denoted by $n_h$ and is a parameter that can be chosen freely. For the neurons in the hidden layer, we use a rectified linear unit (ReLU) activation function, defined as $\text{ReLU}(x) = \max (0,x)$, while the output neurons use an identity activation function.

We note that, apart from the connections that are visible in Fig.~\ref{fig:nn}, each node has also has a \emph{bias} input, which we have omitted from the figure for simplicity. Thus, the total number of (real-valued) weights that need to be estimated is
\begin{align}
	n_w	& = (2M+2L+1)n_h + 2(n_h+1).
\end{align}
Moreover, the linear cancellation stage that precedes the neural network has $2(M+L)$ real parameters that need to be estimated. Thus, the total number of learnable parameters for our proposed neural network canceler is
\begin{align}
	n_{\text{NN}}	& = n_w + 2(M+L).
\end{align}

\section{Computational Complexity}\label{ref:compl}
In this section, we analyze the computational complexity of the polynomial and the neural network canceler in terms of the required number of real additions and multiplications for the inference step (i.e., after training has been performed). We note that, the computational complexity of the training phase is also an important aspect that should be considered, but it is beyond the scope of this paper due to space limitations.

\subsection{Polynomial Canceler}
In order to derive the computational complexity of the polynomial canceler, we ignore the terms in~\eqref{eq:final} for $p=1$ and $q=1$, since these correspond to the linear cancellation which is also performed verbatim for the neural network canceler. This means that there remain $n_{\text{poly}}-M-L$ complex parameters in~\eqref{eq:final}. Moreover, in order to perform a best-case complexity analysis for the polynomial canceler, we assume that the calculation of the basis functions in~\eqref{eq:final} comes at no computational cost. For the non-linear part of~\eqref{eq:final}, $n_{\text{poly}}-M-L$ complex parameters need to be summed, so the minimum required total number of real additions is
\begin{align}
	n_{\text{ADD,poly}}	& = 2(n_{\text{poly}}-M-L-1).
\end{align}

Moreover, assuming that each complex multiplication is implemented optimally using three real multiplications, the number of real multiplications between the complex parameters $h_{p,q}(m)$ and the complex basis functions is
\begin{align}
	n_{\text{MUL,poly}}	& = 3(n_{\text{poly}}-M-L).
\end{align}

\subsection{Neural Network Canceler}
For each of the $n_h$ hidden neurons, $2M+2L+1$ incoming real values need to be summed, which requires a total of at least $(2M+2L)n_h$ real additions. Moreover, at each of the two output neurons, $n_h+1$ real values need to be summed, which requires a total of at least $2n_h$ real additions. The computation of each of the $n_h$ ReLU activation functions requires one multiplexer (and one comparator with zero, which can be trivially implemented by looking at the MSB). Assuming a worst case where a multiplexer has the same complexity as an addition, the total number of real additions required by the neural network canceler is
\begin{align}
	n_{\text{ADD,NN}}	& = (2M+2L+3)n_h.
\end{align}
Excluding the biases which are not involved in multiplications, there are $(2M+2L)n_h$ real weights that are multiplied with the real input values and $2n_h$ real weights that are multiplied with the real output values from the hidden nodes. Thus, the total number of real multiplications required by the neural network canceler is
\begin{align}
	n_{\text{MUL,NN}}	& = (2M+2L+2)n_h.
\end{align}

\section{Experimental Results}\label{sec:results}
In this section, we first briefly describe our experimental setup and then we present results to compare the digital cancellation achieved by the standard polynomial non-linear cancellation method and our proposed neural network based method. We note that all results are obtained using actual measured baseband samples and not simulated waveforms.

\subsection{Experimental Setup}

\emph{Full-Duplex Testbed:}
Our full-duplex hardware testbed, which is described in more detail in~\cite{Balatsoukas2013,Belanovic2013,Balatsoukas2015}, uses a National Instruments FlexRIO device and two FlexRIO 5791R RF transceiver modules. We use a QPSK-modulated OFDM signal with a passband bandwidth of $10$~MHz and $N_c = 1024$ carriers. We sample the signal with a sampling frequency of $20$~MHz so that we can also observe the signal side-lobes. Each transmitted OFDM frame consists of approximately $20,000$ baseband samples, out of which 90\% are used for training and the remaining 10\% are used to calculate the achieved SI cancellation, both for the polynomial model and for the neural network. We use an average transmit power of $10$~dBm and our two-antenna FD testbed setup provides a passive analog suppression of $53$~dB. We note that we do not perform active analog cancellation as, for the results presented in this paper, the achieved passive suppression is sufficient.

\emph{Polynomial Model:}
For the polynomial model, we present results for $M+L = 13$ taps for the equivalent SI channel and for a maximum non-linearity order of $P=7$, since further increasing these parameters results in very limited gains in the achieved SI suppression, and after some point even decreased performance on the test frames due to overfitting. The total number of complex parameters $h_{p,q}(m)$ is $n_{\text{poly}} = 260$, meaning that a total of $2n_{\text{poly}} = 520$ real parameters have to be estimated. As in~\cite{Anttila2014,Korpi2017}, we use a standard least-squares formulation in order to compute all $\hat{h}_{p,q}(m)$.

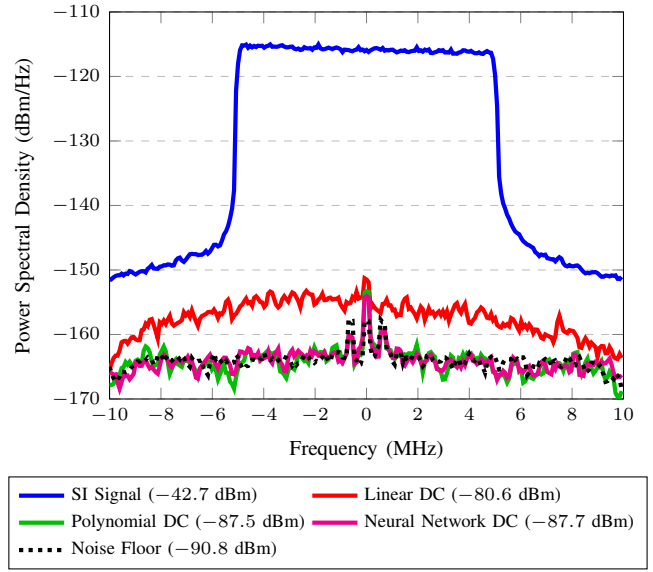
\begin{figure}[t]
	\centering
	\begin{tikzpicture}

	\pgfplotsset{grid style={dashed}}
	
	\begin{axis}[
		width = 0.95\columnwidth,
		height = 0.76\columnwidth,
		xlabel = {Frequency (MHz)},
		ylabel = {Power Spectral Density (dBm/Hz)},
		ylabel near ticks,
		xlabel near ticks,
		xtick distance=2,
		ytick distance=10,
		label style={font=\footnotesize},
		tick label style={font=\scriptsize},
		xmin = -10, xmax = 10,
		ymin = -170, ymax = -110,
		ymajorgrids,
		legend style={at={(0.425,-0.2)},anchor=north,font=\scriptsize},
		legend columns=2,
		legend cell align=left,
		legend entries={SI Signal ($-42.7$ dBm), Linear DC ($-80.6$ dBm), Polynomial DC ($-87.5$ dBm), Neural Network DC ($-87.7$ dBm), Noise Floor ($-90.8$ dBm)}
	]		

		\addplot[blue, ultra thick, solid, each nth point=16, filter discard warning=false] table[x index=0, y index = 1] {figures/data/NLyTestFFT.dat};
		\addplot[red, ultra thick, solid, each nth point=16, filter discard warning=false] table[x index=0, y index = 1] {figures/data/NLyTestLinCancFFT.dat};
		\addplot[green!75!black, ultra thick, solid, each nth point=16, filter discard warning=false] table[x index=0, y index = 1] {figures/data/NLyTestNonLinCancFFT.dat};
		\addplot[magenta, ultra thick, solid, each nth point=16, filter discard warning=false] table[x index=0, y index = 1] {figures/data/NNyTestNonLinCancFFT.dat};
		\addplot[black, ultra thick, dotted, each nth point=16, filter discard warning=false] table[x index=0, y index = 1] {figures/data/NLnoiseFFT.dat};		

	\end{axis}

\end{tikzpicture}%
	\caption{Power spectral densities of the SI signal, the SI signal after linear cancellation, as well as the SI signal after non-linear cancellation using both the polynomial model and the proposed neural network. We also show the measured noise floor for reference.}\label{fig:cancellation}
	\vspace{-0.2cm}
\end{figure}

\begin{figure}[t]
	\centering
	\begin{tikzpicture}

	\pgfplotsset{grid style={dashed}}
	
	\begin{axis}[
		width = 0.95\columnwidth,
		height = 0.76\columnwidth,
		xlabel = {Training Epoch},
		ylabel = {Non-Linear SI Cancellation (dB)},
		ylabel near ticks,
		xlabel near ticks,
		xtick distance=2,
		ytick distance=1,
		label style={font=\footnotesize},
		tick label style={font=\footnotesize},
		xmin = 0.98, xmax = 20,
		ymin = 0, ymax = 8,
		ymajorgrids,
		legend pos = south east,
		legend style={font=\scriptsize},
		legend columns=2,
		legend cell align=left,
		legend entries={Training Frames, Test Frames}
	]		

		\addplot[blue, ultra thick, solid, mark=*] table[x index = 0, y index = 1] {figures/data/convergence_20.dat};
		\addplot[red, ultra thick, solid, mark=square*] table[x index = 0, y index = 2] {figures/data/convergence_20.dat};
    \coordinate (insetPosition) at (rel axis cs:1,0.155);

	\end{axis}

	\begin{axis}[
		at={(insetPosition)},
		anchor={outer south east},
		tiny,
		xlabel = {Training Epoch},
		ylabel = {Non-Linear SI Cancellation (dB)},
		ylabel near ticks,
		xlabel near ticks,
		label style={font=\tiny},
		axis background/.style={fill=white},
		xtick distance=200,
		ytick distance=1,
		xmin = 0, xmax = 1000,
		ymin = 0.95, ymax = 8,
		x tick label style={/pgf/number format/.cd,%
          scaled x ticks = false,
          set thousands separator={},
          fixed}%
		]
    \addplot[blue, ultra thick, solid] table[x index = 0, y index = 1] {figures/data/convergence_1000.dat};
		\addplot[red, ultra thick, solid] table[x index = 0, y index = 2] {figures/data/convergence_1000.dat};
   \end{axis}

\end{tikzpicture}%
	\caption{Achieved non-linear SI cancellation on the training frames and the test frames as a function of the number of training epochs.}\label{fig:conv}
	\vspace{-0.2cm}
\end{figure}
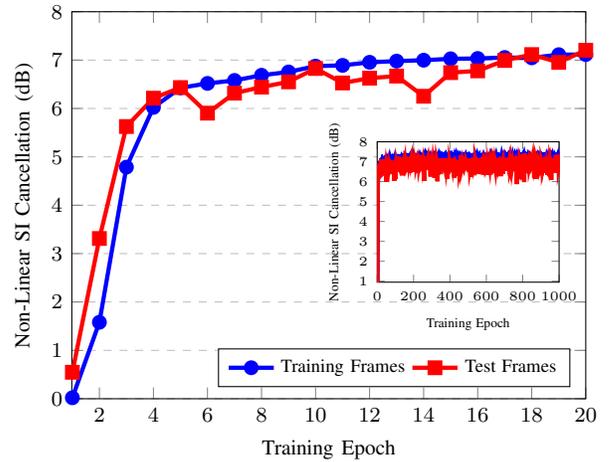

\emph{Neural Network:}
The neural network was implemented using the Keras framework with a TensorFlow backend. Moreover, we use the Adam optimization algorithm for training with a mean-squared error cost function, a learning rate of $\lambda = 0.004$, and a mini-batch size of $B = 32$. All remaining parameters have their default values. In order to provide a fair comparison with the polynomial model, we use $2(M+L) = 26$ input units and $n_h = 17$ hidden units so that $n_w = 495$ weights have to be learned by the neural network and the total number of weights and real parameters that need to be estimated is $n_{\text{NN}} = 521$.

\subsection{Experimental Self-Interference Cancellation Results}
In Fig.~\ref{fig:cancellation} we present SI cancellation results using the polynomial model of Section~\ref{sec:polynomial} and our proposed neural network. We can observe that digital linear cancellation provides approximately $38$~dB of cancellation, while both non-linear cancelers can further decrease the SI signal power by approximately $7$~dB, bringing it very close to the receiver noise floor. The residual SI power for both cancelers is slightly above the noise floor, but this is mainly due to the peaks close to the DC frequency, for which we currently do not have a consistent explanation, and not due to an actual residual signal.

In Fig.~\ref{fig:conv} we observe that after only $4$ training epochs the neural network can already achieve a non-linear SI cancellation of over $6$~dB on both the training and the test frames. After $20$ training epochs the non-linear SI cancellation reaches approximately $7$~dB, which is the same level of cancellation that the polynomial model can achieve, and there is no obvious indication of overfitting since the SI cancellation on the training and on the test data is very similar. Moreover, in the inset figure we observe that allowing for significantly more training epochs does not improve the performance further.

\subsection{Computational Complexity}
For $P=7$, $M+L = 13$, and $n_h = 17$, the polynomial non-linear canceler requires $n_{\text{ADD,poly}} = 492$ real additions and the neural network non-linear canceler requires $n_{\text{ADD,NN}} = 493$ real additions, which is practically identical. However, the polynomial canceler requires $n_{\text{MUL,poly}} = 741$ real multiplications while the neural network canceler only requires $n_{\text{MUL,NN}} = 476$ real multiplications, which is a reduction of approximately $36$\%. We note that, in reality the reduction is much more significant since the calculation of the basis functions in~\eqref{eq:final} also requires a large number of real multiplications.

\section{Conclusion}
In this paper, we have demonstrated through experimental measurements that a small feed-forward neural network with a single hidden layer containing $n_h = 17$ hidden nodes, a ReLU activation function, and $20$ training epochs can achieve the same non-linear digital cancellation performance as a polynomial-based non-linear canceler with a maximum non-linearity order of $P=7$ while at the same time requiring $36\%$ fewer real multiplications to be implemented.

\section*{Acknowledgment}
The author gratefully acknowledges the support of NVIDIA Corporation with the donation of the Titan Xp GPU used for this research. The author would also like to thank Mr. Orion Afisiadis for his aid in carrying out the full-duplex testbed measurements and Prof. Andreas Burg for useful discussions.

\bibliographystyle{IEEEtran}
\balance
\bibliography{spawc2018}

\end{document}